\documentstyle [11pt]{article}

\newcommand{\EqRef}[1]{(\ref{eqn:#1})}
\newcommand{\csec}[1]{\begin{centering}
\newcommand{\tr}[0]{\mbox{tr}}
\newcommand{\im}[0]{\mbox{Im}}

\vspace{1.0cm}

{\Large\bf #1}

\vspace{0.5cm}\end{centering}}

\newcommand{\csub}[1]{\begin{centering}

\vspace{0.5cm}

{\large\em #1}

\vspace{0.25cm}\end{centering}}

\newcommand{\FigCap}[2]{
\ \\
   \noindent
   Figure~#1:~#2
\\
   }

\begin{document}
\title{\bf The role of singularities in chaotic spectroscopy\protect\footnote{
Invited paper for the special issue on
             {\em Chaos and Quantum Transport  in  Mesoscopic Cosmos},
to appear in  Chaos, Solitons \& Fractals (Pergamon/Elsevier)
}}
\author{Per Dahlqvist \\
Mechanics Department \\
Royal Institute of Technology, S-100 44 Stockholm, Sweden\\[0.5cm]
%
}
\date{}
\maketitle

\begin{abstract}
We review the status of the semiclassical trace formula 
with emphasis on the particular types of
singularities that
occur in the Gutzwiller-Voros zeta function for bound chaotic systems.
To understand the problem better we 
extend the discussion to include various classical 
zeta functions and we contrast
properties of axiom-A scattering systems with those of typical bound systems.
Singularities in classical zeta functions contain topological and
dynamical information, concerning e.g.
anomalous diffusion, phase transitions among generalized Lyapunov exponents,
power law decay of correlations.
Singularities in semiclassical zeta functions are artifacts
and enters because one neglects some quantum
effects when deriving them, typically by
making saddle point approximation when the saddle points are not
enough separated. 
The discussion is exemplified by the Sinai billiard
where intermittent orbits associated with neutral orbits induce
a branch point in the zeta functions. 
This singularity is responsible for a diverging diffusion constant
in Lorentz gases with unbounded horizon.
In the semiclassical case there is
interference between neutral orbits and intermittent orbits.
The Gutzwiller-Voros zeta function 
exhibit a branch point
because it does not take this effect
into account. Another consequence is that individual states, high up in
the spectrum, cannot be resolved by 
Berry-Keating technique.

\end{abstract}

\csec{I. Introduction}

Research in quantum chaos has to a large extent been focused on conceptual
questions such as 
{\em how is classical chaos revealed in the $\hbar \rightarrow 0$
limit of quantum mechanics} and {\em what are the properties of quantum
spectra of classically chaotic systems}.
During the last decade another hot topic has been periodic orbit quantization. 
Model calculation of quantum spectra using periodic orbit techniques
are greatly simplified if one restrict oneself to
systems with two degrees of freedom.
Applications to three dimensional systems are then restricted to cases
where some internal symmetry reduces the system by one degree of freedom.
One example is the diamagnetic Kepler problem, 
see e.g. \cite{THM95} and references therein.
The simplest conceivable two dimensional systems are billiards.
Billiards are of great use 
in theoretical work on semiclassical mechanics.
But billiard like systems, of real interest, and
 of any imaginable shape can now be constructed
in the laboratory, there are several examples in the present volume. 
So the stage is set for a fruitful interplay
between quantum chaos and mesoscopic physics.

To make a successful semiclassical computation
one needs to control the semiclassical limit of the Green function.
One can then compute anything from conductance in 
mesoscopic devices \cite{BarrLA}
to wave functions.
Another  information contained in the Green function is the quantum spectrum,
which can be uncovered by taking its trace.
In bound systems the Green functin can be expressed in 
terms of the eigenfunctions and
eigenvalues of the Hamiltonian
\begin{equation}
G(x,x',E)=\sum_{i} \frac{\Psi_i^*(x)\Psi_i(x')}{E-E_i} \ \ ,
\end{equation}
and the trace reads
\begin{equation}
tr G(x,x',E)= \int G(x,x,E) dx=\sum_{i} \frac{1}{E-E_i} 
\end{equation}
whose semiclassical limit for chaotic system was derived by Gutzwiller
\cite{Gut90}
\begin{equation}
      tr G(E)=g_{o}(E)
           +\frac{1}{i{\hbar}}\sum_{p}T_{p}\sum_{n=1}^{\infty}
            \frac{1}{|M_{p}^{n}-I|^{\frac{1}{2}}}
            \,e^{in
               \left[S_{p}/\hbar-\mu_{p}\frac{\pi}{2}\right]}
            \ \ .
\end{equation}
The index $p$ labels the primitive periodic
orbits, $S_{p}$ is the action integral along the orbit, $T_{p}$
$(=\partial S_{p}/\partial E)$ its period,
$M_{p}$ is the linearized Poincar\'{e} map around the orbit and $\mu_{p}$ is a
phase index.
Finally, $g_{o}(E)$ provides the mean level distribution, which may be
interpreted as the contribution from classical orbits of zero length.

The trace formula can be
recasted to the Gutzwiller-Voros {\em zeta function}\cite{Vor88} 
\begin{equation}
     Z_{GV}(E)=\prod_{p}\prod_{m=0}^{\infty}
          \left(1-\frac{e^{i\left[S_{p}/\hbar-\mu_{p}\frac{\pi}{2}\right]}}
            {|\Lambda_{p}|^{1/2}\Lambda_{p}^{m}}\right) \ \ .  
\label{eqn:GVzeta}
\end{equation}
The oscillating part of the trace formula is given by the logarithmic derivative
of the zeta function.
The only continous time systems we are going to consider are billiards.
The action $S_p$ is then related to the geometric length $L_p$ of the orbit
as $S_p=L_pk$ where $k$ is the momentum. Usually we are going to consider zeta
functions as functions of the complex variable $s=ik$.
This paper is devoted to the study of classical and 
semiclassical zeta functions.
In particular, we will be interested in the presence of various kinds of 
singularities.
This might seem a highly technical problem but by putting this theme
in focus we will shine some new light on
many of the problems that are associated with
semiclassical quantization. The paper is divided into two parts.
First a nonechnical review of the subject. Then follows a series of 
more technical
discussions to illustrate various aspects of the preceding story.

\newpage

\csec{II. A Story - The rise and fall\\ 
of periodic orbit theories in quantum mechanics}

Our story will begin with Berry's 'tour de force' paper\cite{Berr81}
{\em Quantizing a Classically Ergodic System, Sinai's
Billiard and the KKR method}.
Centered around a numerical method Berry discusses
virtually all issues now known under the phrase of {\em Quantum Chaos}, like
e.g. level statistics and periodic orbit expansions.
Taking the semiclassical limit of level density he obtains the Gutzwiller
trace formula for the unstable orbits and the Berry-Tabor like expression
for the neutral (or marginally stable) orbits.

At the time it was premature to
start from the trace formula and compute (semiclassical approximations of) the
eigenvalues. Berry's numerical scheme was successful because of a major trick;
a transformation of the badly converging sum over lattice points 
to a highly convergent sum
over points in the dual lattice. 
One of the major problem with periodic orbit expressions is to make the
corresponding trick for the Gutzwiller sum.
We will discuss this problem from several points of view in the following.

Later, Gaspard and Rice \cite{GR89} were inspired by Berry's method to 
construct a numerical
scheme for determining scattering resonances in the 3-disk scatterer,
and similar systems. 
From one point of view this is an easier problem
(from another, more conceptual it is not), 
the disk lattice is finite and there is no need to
go over to the dual lattice.

The 3-disk is easy to deal with also from a semiclassical point of view, 
as was realized by
Cvitanovi\'{c} and Eckhardt \cite{CE93}. The Gutzwiller-Voros {\em zeta
function}\cite{Vor88} closely resembles zeta functions studied in 
classical problems.  The
3-disk problem is ideal for 
application of classical zeta functions. 
It has a simple topology, provided
the disks are enough separated, the dynamics may be mapped 
to a dynamics among a finite set of symbols. This is known as 
{\em symbolic dynamics}.
The open 3-disk is also hyperbolic; the stabilities of periodic orbits are
exponentially bounded with their length. 
A system with these two properties is said to be of Axiom-A type.
This property gives nice convergence to the 
Dirichlet series obtained by expanding the Euler product representation of the
zeta function \EqRef{GVzeta}, 
this is usually called a 
{\em cycle
expansion} \cite{AAC90,CE93}.
The classical resonances as well as the quantum
resonances can thus be efficiently computed by a
moderate number of periodic orbits.

Still, the situation is not ideal. The Gutzwiller-Voros zeta function 
does exhibit
poles \cite{ER93} which limits convergence 
of cycle expansions far down in the complex $k$-plane.
A theorem of Rugh\cite{Rugh} says that 
a certain type of zeta functions, Fredholm determinants
of evolution operators with multiplicative weights, are entire
for Axiom-A systems.
A Fredholm determinant may look something like
\begin{equation}
F(z)=\prod_p \prod_{m=0}^{\infty} (1-\frac{w_p e^{-s L_p}}{\Lambda^m})^{m+1}
\end{equation}
The Gutzwiller-Voros zeta function lacks the {\em right} structure among the
higher $m$ factors to make it entire.
There is a lot of interesting work \cite{CR93,CVW96} on improving the 
Gutzwiller-Voros zeta
function by manipulating these higher factors.
The by far most interesting is the {\em quasi classical} theory of
Cvitanovi\'{c} and Vattay \cite{CVW96}.
Starting from the hydrodynamical formulation of Quantum Mechanics of Madelung
they construct a multiplicative evolution operator whose Fredholm 
determinant should contain
the semiclassical eigenvalues in its spectrum.
The derivation
 avoids the multitude of stationary phase approximations
in the standard derivation of the trace formula.
Still, the quasiclassical Fredholm determinant contains extra zeros
which have nothing to do with the quantum spectrum, and they seriously slow 
down convergence.
So the surprising thing is that, 
if interpreted as an asymptotic series, the Gutzwiller-Voros
zeta function is in many respects still superior
to its competitors \cite{Wir95}.

Due to the simplicity of the 3-disk scatterer it is an ideal testing ground
for various corrections to the Gutzwiller-Voros zeta function, like
diffraction-corrections \cite{VWR95,RVW} and $\hbar$ corrections\cite{VR96}.
Note that the circumstances that make the 3-disk classically and 
semiclassically {\em nice}
also make diffraction corrections {\em small}; semiclassical 
formulas fail if the point
particle scatter in an extreme forward angle, or if it passes extremely close
to a disk and  this is never the case in the (enough separated) 3-disk.
The main diffraction effect for this system is then due to creeping 
and this is indeed a
small effect.

When including diffraction effects in axiom-A systems one can say 
that the classical
phase space is enlarged to include diffractive orbits associated with some
discontinuity such as a point or wedge or circular disk. 
This can be done without risking that
stationary points come to close. We will see that this is very different 
from the
problem intermittency
is going to present us with in bound systems.

We have dwelt on Axiom-A scattering 
systems at some length, mostly to appreciate
how different from bound systems they are.
When we now are turning our interest towards bound systems
we will encounter singularities that have
nothing to do with the higher
$m$ factors in Fredholm determinants, 
so we will frequently omit them. Their subtle role in bound systems
is far beyond the scope of this paper.

If the open 3-disk system was an archetype for open chaotic scattering systems
the Sinai billiard is a suitable archetype for bound chaotic systems.
It is in many ways generic (see below) but is simple enough to 
allow some analysis.
This system is either represented as a circular scatterer inside a square or
as an infinite lattice of disks, cf. fig 1. 
The latter representation is an example of a {\em Lorentz
gas}. Lorentz gases are among the simplest conceivable systems exhibiting 
statistical
behaviour such as diffusion \cite{PDlyap}. 

Bounded system are generally not of Axiom-A type.
We will now discuss various deviations and how they affect the
analyticity of the zeta functions.

The topology is generally extremely complicated.
Given a bound system, there is a priori no reason to believe that there should
exist a simple symbolic dynamics. Often one can find a symbolic coding but
the grammar rules are complex, in the sense
that the number of forbidden substrings increases exponentially with length.
Mathematically, such a symbolic dynamics is known as an 
infinite subshift.
The basic mechanisms for {\em pruning} of the symbolic dynamics
can be very different.
In dispersive billiards pruning enters because a
trajectory from one disk to another
can be obstructed by an intermediate disk. 

Very little is known about how pruning affects
the analyticity of zeta functions. In ref, \cite{PDgap} we propose that
the lack of
finite symbolic dynamics may induce a natural boundary in the complex plane
through which analytic continuation is not (even in principle) possible.
We discuss this in little more detail in section IIIa.

Intermittency is another abundant property of bound
chaotic systems.
We will consider a strong form of intermittency associated
with the presence of
neutral orbits in the system.
Orbits accumulating towards this neutral orbit will have Lyapunov exponents
tending to zero.
Many ergodic billiards, such as the stadium\cite{Tann96} and the Sinai billiard
have this property. 
The effect is relevant for Hamiltonian systems with mixed
phase space because the boundaries of stable islands
are neutrally stable and
will be surrounded by intermittent orbits.
So the neutral orbit in a chaotic billiard may be looked upon as a 
stable island of
zero area\cite{Tann96}.
Strong intermittency appears to
imply a branch point singularity \cite{PDsin,PDLA} at the origin (s=0)
as discussed in section IIIb.
It is possible to construct bounded billiards without neutral orbits.
One example of is (the unit cell) of a Lorentz gas on a triangular lattice,
 provided the disks are large enough, see \cite{CGS92}.
Such systems often exhibit some weaker form of intermittency.

In classical zeta functions,
these singularities 
play a dual role. They hamper convergence of cycle expansions seriously.
But they are also important
carriers of dynamical and topological information
such as power law decay of correlations, 
anomalous diffusion and phase transitions among
the generalized Lyapunov exponents\cite{PDlyap},
and should not be considered as undesired mathematical obstacles. 
In section IIIc we, as an example, discuss the relation between 
the singularity of
the zeta function and
anomalous diffusion in the Sinai billiard.

Singularities will be present in virtually any type of
zeta function, and in particular the Gutzwiller-Voros zeta function 
(see section IIIb).
This zeta function is supposed to approximate the spectral 
determinant $\Delta(k)$
(apart from the smooth Weyl factor).
This function obeys a functional equation $\Delta(k)=\Delta(-k)$
and the singularities we have discussed violate
this functional equation.
In the Gutzwiller-Voros zeta function they are artifacts
and enter because
one neglects some quantum
effects  when deriving it,
like making saddle point approximations when the saddle points are not
enough separated. This problem is intimately associated with intermittency.
In section IIIe 
we demonstrate that the problem with interfering stationary points
actually occur in the Sinai billiard and this
is what lets a branch point enter the zeta function.

If the system under study is hyperbolic the situation is somewhat better, but
one has still to bother about pruning.
The main pruning mechanism in dispersive billiards is,
as we said, due to trajectories between two disks 
being obstructed by a third disk. Orbits
that are nearly pruned are therefore greatly affected 
by diffraction effects so pruning and
diffraction goes hand in hand and must be dealt with in a coherent way, 
let's call such a
hypothetical scheme {\em Quantum pruning}.
Pruning in time reversible system can often be described by a 
{\em pruning front}
\cite{KaiThesis}.
Intuitively we can say that the quantum counterpart 
is not sharp as the classical one.
Recall that we expect that a generic pruning front may induce
a natural boundary
in classical zeta functions, so the fuzziness in the quantum counterpart should
prevent such a singularity to occur.

How serious the problems of the Gutzwiller-Voros zeta functions are depend
on the method by which one computes the zeros.
There are mainly two approaches. 
The first is the cycle expansion approach, previously
discussed, used mainly in open Axiom A systems.
This relies heavily on strong convergence properties of the zeta functions.
An alternative for bound systems
is to use the method of Berry-Keating\cite{BK90}. 
The main idea is the following. 
The existence of a functional equation for the spectral determinant
imply that one only needs to know
the length spectrum (the fourier transform of the level density) up to
half the Heisenberg length\footnote{the length corresponding to one 
mean level spacing} in order to compute the spectrum.
The most accurate computation of spectra of the Gutzwiller-Voros
zeta functions for bound systems
has been performed using this
approach\cite{SS91,Tann91}.
The advantage of the method is that one only needs a finite number of periodic
orbits to compute a finite number of levels, and one only needs to scan the
real energy axis.
The approach of Berry-Keating assumes that the spectral determinant may 
be expressed
as a Dirichlet series.

There are now growing evidence \cite{Prim95,Tann96} that even the 
Berry-Keating method
will eventually
break down for billiards.
In ref. \cite{Prim95} the authors argue that 
the majority of orbits in the Sinai billiard
with length $L>O(k^{2/3})$ is affected by
diffraction and that
these corrections are of the same order as the original contribution.
Recalling that the Heisenberg length in billiards scales as $L_H =O(k)$ this
implies that the Berry-Keating scheme will break down as
$k \rightarrow \infty$,
provided that diffraction
effects cannot be accounted for.
The occurrence of a branch point singularity is actually just another 
side of this
coin. It indicates that the tail of the Dirichlet series suffers so 
badly from intermittency
and the nearby stationary points so as to make the series divergent
(a branch point at $k=0$ makes the real $k$-axis 
where we want to find our eigenvalues
the border of convergence).
The entanglement
between neutral and intermittent orbits implies that the 
representation of the spectral
determinant in terms of a Dirichlet series is lost.
There is thus no obvious implementation of the Berry-Keating method. 

The problem with interfering saddle points has nothing to do with 
the singular nature of billiards.
In smooth chaotic potentials it is even harder to realize a complete 
symbolic dynamics.
The abundance of pruning is deeply associated with the existence of small
stability islands. These islands will immediately cause problems with
intermittency and close saddle points.

It should be stressed that such a failure of predicting
individual eigenstates does not render the trace formula useless.
One may take two approaches.
One is to take the trace formula more or less
as it stands and ask for the limitations of it. 
What information can be gained from this simple and beautiful formula?
To use it as a theoretical tool one also needs a better understanding of the
asymptotic properties of periodic orbits \cite{PDsin,PDLA} far beyond the
Ozorio-Hannay sum rule \cite{Ozo}.
The
other approach is to try to improve the trace formula in various ways. 
The two approaches are well motivated and
are of course not independent of each other.
It is an intellectual challenge to improve periodic orbit theories
to predict eigenstates higher and higher but it is perhaps not 
always what applications
ask for.

Among all systems exhibiting hard chaos, 
systems with neutral orbits are among the worst,
without neutral orbits the situation would probably be better.
The conceptual problems 
of systems with neutral orbits
are similar to those of generic systems with
mixed phase space, and are thus of fundamental importance.
The relative simplicity of the Sinai billiard make it an ideal model system 
for addressing these issues.

\csec{III. Five Illustrations}

We will now illustrate the story in the previous section with a series of more
technical discussions.
We are trying to understand the sense of singularities in zeta functions
in the classical case and the nonsense of them in the semiclassical case.
We will start our expos\'{e} by considering 
classical zeta functions for discrete time systems (i.e. {\em maps}),
move on to continous time systems (in particular the Sinai billiard)
and finally discuss semiclassical zeta functions.

\csub{a. Symbolic dynamics in bound chaotic systems}

We begin by considering one dimensional maps
$x \mapsto f(x)$.
Much dynamic information, like  different kind of entropies and
generalized Lyapunov exponents, is encoded in 
the following one-parameter
family of zeta functions \cite{Beck}
\begin{equation}
1/\zeta(\beta,z)=\prod_p (1-\frac{z^{n_p}}{|\Lambda_p|^{1-\beta}})  \  \ .
\label{eqn:dyndef}
\end{equation}
The product in \EqRef{dyndef}
runs over all primitive periodic orbits $p$, having period
$n_p$ and
stability $\Lambda_p=\frac{df^{n_p}}{dx}|_{x=x_p}$ with $x_p$ being any
point along $p$.
If $\beta=1$ no metric information enters the zeta
which is now called a {\em topological zeta function}
\begin{equation}
1/\zeta_{top} (z) =\prod_p (1-z^{n_p}) \  \ .
\end{equation}
The leading zero  $z_0$ of $1/\zeta_{top} (z)$ 
(the one with smallest modulus) and the
topological entropy $h$ are related by $h=-\log z_0$.

We will restrict our attention to the topological zeta function for 
unimodal (one-humped)
maps with one external control parameter $f_\lambda(x)=\lambda g(x)$.
Symbolic dynamics is introduced by mapping a time series 
$\ldots x_{i-1}x_i x_{i+1} \ldots$
onto a sequence of symbols $\ldots s_{i-1}s_i s_{i+1} \ldots$
where 
\begin{eqnarray}
s_i=L & x_i < x_c\\
s_i=C & x_i=x_c  \  \ , \\
s_i=R & x_i>x_c
\end{eqnarray}
and $x_c$ is the critical point of the map (i.e. maximum of $g$).
The kneading sequence is the itenary of the critical point.
The allowed symbol sequences can be determined from it.
All unimodal maps (obeying some further constraints)
with the same kneading sequence have the same set of 
periodic orbits\cite{CE80}.

If $\beta\neq 1$ in \EqRef{dyndef} the 
individual details of $f(x)$ is reflected
in the zeta function, but for
{\em tent map} 
\begin{equation}
x \mapsto f(x)= \left\{ \begin{array}{ll}
\lambda \cdot x & x \in [0,1/2] \\
\lambda \cdot (1-x) & x \in (1/2,1] \end{array} \right.    \label{eqn:tent}
\  \ .
\end{equation}
(where the parameter $\lambda \in [0,2]$)
the general zeta function \EqRef{dyndef} is obtained from $1/\zeta_{top}$
by simple rescaling,
\begin{equation}
1/\zeta(\beta,z)=1/\zeta_{top} (z/\lambda^{1-\beta})  \  \  . 
\label{eqn:Resc} \end{equation}
The topological entropy is $h=-\log\lambda $.

The set of periodic points of the tent map is countable. A consequence of 
this fact is that the set of parameter values for which the kneading sequence
is periodic or eventually periodic ({\em preperiodic})
are countable and thus of measure zero and
consequently {\em the kneading sequence is aperiodic for almost all $\lambda$}.
For general unimodal maps the corresponding statement is that
the kneading sequence is aperiodic for almost all topological entropies.

For a given periodic kneading sequence of period $n$,
$\underline{I}_\lambda =\overline{PC}=\overline{s_1 s_2 \ldots s_{n-1}C}$
there is a simple expansion for the topological
zeta function. 
Now let $a_i=1$
if $s_i=L$, and  $a_i=-1$ if $s_i=R$.
Then the expanded zeta function is a polynomial
of degree $n$ 
\begin{equation}
1/\zeta_{top} (z)=\prod_p (1-z^n_p)=(1-z)\cdot \sum_{i=0}^{n-1}b_i z^i
\  \ ,   \label{eqn:simpa}
\end{equation}
where
\begin{equation}
b_n=\prod_{i=1}^{n}a_i \ \ .   \label{eqn:simpb}
\end{equation}
Aperiodic and preperiodic
kneading sequences is accounted for by simply replacing
$n$ by $\infty$.
An important consequence of \EqRef{simpa} and \EqRef{simpb} is that the 
sequence 
$\{ b_i \}$ has a periodic tail if and only if the
kneading sequence has
one (their period may differ by a factor of two though).

The analytic structure of the function represented by the series
$\sum b_i z_i$ depends on whether the tail of $\{ b_i \}$ is periodic or not.
If the period of the tail is $N$ we can write
\begin{equation}
1/\zeta_{top}(z)=p(z)+q(z)(1+z^N+z^{2N}\ldots)=p(z)+\frac{q(z)}{1-z^N}  \ \ ,
\end{equation}
for some polynomials $p(z)$ and $q(z)$.
The result is a set of poles spread out along the unit circle.
This applies to the preperiodic case.
An aperiodic sequence of coefficients would formally correspond to an 
infinite $N$
and it is natural to assume that the singularities will fill the unit
circle. 
There is indeed a theorem ensuring that this in a sense is the case
(provided the $b_i$'s can only take an finite number of values).
The unit circle becomes a
natural boundary.
A function with a natural boundary lacks an analytic continuation outside
it.


What happens with $1/\zeta(\beta,z)$ for $\beta\neq 1$ 
if we make our unimodal map nonlinear 
but still hyperbolic (that is $|f'|\geq C>1)$?
All that is known is that the the zeta function is analytic 
inside a certain radius and an obvious guess is that this radius
is limited by a natural boundary in the generic case.
It is natural to expect that the natural boundary moves outward 
 from the
unit circle as $\beta$ is decreased, and probably it gets deformed.

Even less is known if $f$ is intermittent, some results have recently 
been made \cite{Rugh96}.
Except for the branch cut (see section IIIb)
the zeta functions ($\beta=0$)
is holomorphic in a certain region enclosing the unit circle.

If the system is time reversible one expects pruning to be described
by a {\em pruning front} 
which is the two-dimensional generalization of the kneading
sequences\cite{KaiThesis}. 
Moreover, if the system is time continous the expanded zeta function
is a Dirichlet series rather than a power series.
Knowing that natural boundaries are abundant also in Dirichlet series we 
have at least
to focus the possibility of the existence of 
natural boundaries in zeta functions
of systems with no simple symbolic description.

\csub{b. Intermittency}

Intermittency means that the system switches between chaotic behaviour and
quasi-integrable behaviour.
We will consider a
strong form of
intermittency associated
with the presence of (one or more)
neutrally stable orbits. 
To illustrate how this affects
the analyticity of the zeta function it is instructive to consider
a one dimensional
map $x \mapsto f(x)$ on the unit interval, with
\begin{equation}
f(x)=\left\{ 
\begin{array}{cc}
x+2^s x^{1+s} & 0\leq x <1/2\\
2x-1 & 1/2 \leq x \leq 1
\end{array}  \ \ ,
\right.
\end{equation}
where $s> 0$, see fig 2. For $s=0$ the map is just the binary shift map,
which is uniformly hyperbolic, but for
$s>0$ it is intermittent; the fix point $x=0$ is neutrally
stable: $f'(0)=1$.
The map admits a binary coding, 
as before we associate the letter $L$ with the left leg,
and $R$ with the right leg. The neutral fix point now 
corresponds to the periodic orbit
$\overline{L}$.
The symbolic dynamics is by construction complete.

We will consider the zeta function \EqRef{dyndef} with the neutral 
fix point pruned
\begin{equation}
1/\zeta_f(z)=\prod_{p\neq \overline{L}} 
(1-\frac{z^{n_p}}{|\Lambda_p|^{1-\beta}})
\label{eqn:Zdef}  \ \ .
\end{equation}
We now suppress the parameter $\beta$. The new index indicates what map
the zeta function refer to.

A power series representation of $1/\zeta$ is obtained by 
expanding this product:
$1/\zeta_f(z)=\sum_n a_i z^n$.
The nature of the 
leading singularity will be reflected in the asymptotics
of the sequence $\{ a_i \}$.
To get an idea what this asymptotic behaviour may be
we consider the zeta function of closely related map $\hat{f}$, a piecewise
linear approximation of $f$.
We define $\hat{f}$ as a continous function, coinciding with $f$ on a sequence
of points $\hat{f}(c_n)=f(c_n)$ where the $c_n$'s are 
the inverse images of the critical point (cf fig 2)
\begin{eqnarray}
c_0=1/2\\
c_{n+1}=f^{-1}_L (c_n)  \ \ ,
\end{eqnarray}
and linear in the intervals $[c_{n+1},c_n]$.

It is relatively easy (see e.g. \cite{Iso95}) 
to show that the sequence $\{ c_n \}$ has the asymptotic
behaviour
\begin{eqnarray}
c_n \sim n^{-1/s} & n \rightarrow \infty  \ \ .
\end{eqnarray}
The construction of $\hat{f}$ gives it a very simple cycle expansion
\begin{equation}
1/\zeta_{\hat{f}}(z) = 1-\sum_{n=0}^{\infty} 
\frac{z^{n+1}}{\Lambda_{\overline{L^nR}}}\equiv \sum_n \hat{a}_n z^n \ \ .  
\end{equation}
The stabilities $\Lambda_{\overline{L^nR}}$
are simply related to the $c_n$'s
\begin{equation}
\Lambda_{\overline{L^nR}}=2\frac{c_n-c_{n+1}}{c_0-c_1}  \  \  ,
\end{equation}
with the asymptotic behaviour
\begin{equation}
\Lambda_{\overline{L^nR}}\sim n^{(s+1)/s}   \label{eqn:slope}  \  \ .
\end{equation}
The asymptotic behaviour of the coefficients are thus
$\hat{a}_i=O(i^{-(s+1)/s})$ and it seems likely that the same holds
for the sequence $\{a_i \}$.
This suggest that $1/\zeta_f(z)$ contains a singularity of the type
\begin{equation}
\begin{array}{cc}
(1-z)^{\alpha-1}  &  \alpha \not\in N\\
(1-z)^{\alpha-1}\log (1-z) & \alpha \in N^+  \ \ ,
\end{array}  \label{eqn:sing}
\end{equation}
with
\begin{equation}
\alpha (\beta ,s)=\frac{(1-\beta)(s+1)}{s}  \ \ ,
\end{equation}
as can be realized through the Tauberian theorems for power series.

We will now move on to the Sinai billiard.
In billiards with continous time 
one cannot study zeta functions in complex $z$ since the lengths $L_p$ 
are not integer
multiples of some unit length. Instead we 
replace $z$ by
$\exp(-s)$ and formulate the
the zeta functions
\begin{equation}
Z(s)=\prod_p (1-\frac{e^{-sL_p}}{|\Lambda_p|^{1-\beta}})
\end{equation}
where $p$ runs over all primitive unstable cycles.
A semiclassical zeta function is formally obtained as $\beta=1/2$, we
only need to reinsert the Maslov indices (which can be done properly by a 
multiplicative
weight). 

The presence of a simple symbolic dynamics in the previous example 
made it easy for
us since (at least in the piecewise linear approximation) 
we only needed to consider one sequence of periodic orbits accumulating at
the neutral fixpoint. In the Sinai billiard there are
neutral orbits in a (finite) number of directions.
A sequence of periodic orbits accumulating towards a
neutral orbit
is indicated in fig 1, let us denote the sequence $q_n$.
The stability of the cycles obey a power law $\Lambda_{q_n} \sim n^{2}$ 
asymptotically and since
$T_{q_n}\sim n$ the local Lyapunov exponent 
$\log \Lambda_{q_n}/T_{q_n}$
goes to zero and we can again expect
a branch cut. But the order of this singularity cannot be estimated as
easily as in the previous one dimensional map. The reason is the lack
of simple symbolic dynamics. 
There is a simple symbolic coding \cite{PDsin} but it is
heavily pruned for cycles having segments close to the neutral orbits.
We must rely on more elaborate methods to investigate the singularity.

In a series of paper 
\cite{PDreson,PDsin,PDLA,PDsmall,PDlyap,PDzak}
we have investigated this problem within the frame
work of the BER (acronym for Baladi-Eckmann-Ruelle) approximation \cite{BER}.
The basic idea is that for intermittent systems
one can usually find a 
surface of section whose associated 
Poincar\'{e} map is not intermittent, ideally
it is exponentially mixing. An approximate zeta function 
for the flow through this map can be formulated as a 
an average over the surface of section
\begin{equation}
Z_{BER}(s)=1-\langle 
e^{-s\Delta_s(x_s)}|\Lambda(x_s)|^\beta\rangle_{x_s \in sos} 
\label{eqn:Zflow} \ \ ,
\end{equation}
where $\Delta_s(x_s)$ is the traveling length to the next intersection 
with the surface
of section and $\Lambda(x_s)$ its associated instability, both being functions
of the surface of section coordinate $x_s$.

In Hamiltonian chaotic system this average is easily performed since
the invariant density is uniform.
It is convenient to define
\begin{equation}
p_\beta (\Delta)=
\langle \delta(\Delta-\Delta_s(x_s))|\Lambda(x_s)|^\beta\rangle_{sos}
\label{eqn:pDeltadef} \ \ .
\end{equation}
For $\beta=0$ this is just the (probability) distribution of recurrence times
to the surface of section.
The BER zeta function is obtained by a simple Laplace transform
\begin{equation}
Z_{BER}(s)=1-\int_0^\infty e^{-s\Delta} p_{\beta}(\Delta) d\Delta
\end{equation}

In the Sinai billiard the obvious choice of surface of section is the disk.
The function $p_\beta(\Delta)$ of the disk-to-disk map
can be controlled well: 
for $\beta=0$ we get in the limit $R \rightarrow 0$ \cite{PDsmall}
\[
p_0(\Delta)\sim
\]
\begin{equation}
 \left\{ \begin{array}{ll}
\frac{12R}{\pi^2} & \xi<1\\
\frac{6R}{\pi^2 \xi^2}(2\xi+\xi(4-3\xi)\log(\xi)+
4(\xi-1)^2 \log(\xi-1)-(2-\xi)^2\log|2-\xi|)&
\xi>1 \end{array} \label{eqn:psmall} \right.  \ \ ,
\end{equation}
where
$\xi=\Delta 2R$. 

For finite $R$ we only have simple expression for the tail
\begin{eqnarray}
p_\beta(\Delta) \sim \frac{1}{\Delta^{3-3\beta/2}}  \ \ .
\end{eqnarray}
The prefactor can be computed exactly for any disk radius $R$
but this is not our concern here.
It is also well known the the mean is given by
\begin{equation}
\int_0^\infty \Delta p_{\beta=0}(\Delta) d\Delta=\frac{1}{2R}+O(R)  \  \  ,
\end{equation}
an expression we will need in section IIIc.

A power law tail of $p\sim \Delta^{-\alpha}$ thus gives a leading singularity
\begin{equation}
\begin{array}{cc}
s^{\alpha-1}  &  \alpha \not\in N\\
s^{\alpha-1}\log s & \alpha \in N^+  
\end{array}  
\end{equation}
to the BER zeta function.

How accurate the BER approximation is depend on the underlying map.
For the Sinai billiard the disk to disk map is hyperbolic.
It is not known if it i exponentially mixing 
but 
the correlation function 
can at least \cite{Bum} be bounded by a stretched exponential.
We expect that the first few terms in the generalized series
expansion is exactly reproduced by the BER approximation, at least we expect
that \cite{PDzak}
\begin{equation}
Z(s)=Z_{BER}(s)+O(s^2)  \ \ ,
\end{equation}
but higher terms are also described well by the BER approximation \cite{PDsin}.
For $\beta=0$ this would imply that the leading singularity
$s^2 \log s$ is exactly reproduced
by the BER approximation. 
For the semiclassical case $\beta=1/2$ we get a leading singularity
of the form $s^{5/4}$ and thus an algebraic branch point at $s=0$.
The introduction of Maslov indices will affect the size but not the order
of the singularity.

\csub{c. The role of singularities in classical zeta functions}

What are classical zeta functions good for?

The most well known application is the relation between the zeros of
$1/\zeta_{\beta=0}$ and the decay of correlations.
This problem is rather complicated mathematically and 
we will instead consider the somewhat simpler problem
of computing averages 
of chaotic systems \cite{PCLA}. In particular we will be interested
in computing the diffusion constants
but we'll start by considering
more general averages.

We assign a weight $w(x_0,t)$ to the trajectory starting at phase space point
$x_0$ and evolving during time $t$ to point $x(x_0,t)$.
There will be an important technical restriction on $w(x_0,t)$:
it must be multiplicative
along the flow, which means that
 $w(x_0,t_1+t_2)=w(x_0,t_1)w(x(x_0,t_1),t_2)$.
The phase space average of $w(x_0,t)$ may be expanded in terms of periodic
orbits as \cite{PDlyap}
\begin{equation}
\lim_{t \rightarrow \infty}
\langle  w(x_0,t)\rangle  = \lim_{t \rightarrow \infty} 
\sum_p T_p \sum_{r=1}^{\infty} w_p^r \frac{\delta(t-rT_p)}
{|\Lambda_p|^r}  
 \ \ ,
\label{eqn:tracedef}
\end{equation}
where $r$ is the number of repetitions of primitive orbit $p$ and $w_p$ is
the weight integrated along cycle $p$.
This trace may be formulated in terms of
zeta functions in the following way
\begin{equation}
\lim_{t \rightarrow 
\infty}\langle w(x_0,t)\rangle  = \lim_{t \rightarrow \infty} \frac{1}{2\pi i}
\int_{-i\infty+a}^{i\infty+a} e^{st}\frac{Z_w'(s)}{Z_w(s)}ds
\label{eqn:intlogder}   \ \ , 
\end{equation}
with the zeta function \begin{equation}
     Z_w(s)=\prod_{p}
          \left(1-w_p \frac{e^{-sT_{p}}}
    {|\Lambda_{p}|}\right) \ \ .
\label{eqn:zetaw}
\end{equation}

In the previous section we studied the thermodynamic weight 
$w_p=|\Lambda_p|^\beta$ but we will now turn to the problem of 
diffusion of interest
for the present volume.
To this end we follow \cite{CEG94} and
introduce the weight
\begin{equation}
w_{diff}(x_0,t)=e^{\bar{\beta}\cdot(\bar{x}(x_0),t )-\bar{x}_0)}  \ \ ,
\label{eqn:wD}
\end{equation}
where $\bar{x}$ is the configuration space part of the phase space vector $x$.
The diffusion constant may now be expressed in terms of the associated zeta
function\cite{CEG94,PDlyap}
\[
D=\lim_{t \rightarrow \infty}\frac{1}{2t} 
\langle(\bar{x}(x_0),t )-\bar{x}_0)^2\rangle
=\lim_{t \rightarrow \infty}\frac{1}{2t}(\frac{d^2}{d\beta_1^2}
+\frac{d^2}{d\beta_2^2})\langle
e^{\bar{\beta}\cdot(\bar{x}(x_0),t )-\bar{x}_0)}\rangle\mid_{\beta=0} \]
\begin{equation}
= \lim_{t \rightarrow \infty}\frac{1}{2t} \frac{1}{2\pi i}
\int_{-i\infty+\epsilon}^{i\infty+\epsilon} e^{st} 
(\frac{d^2}{d\beta_1^2}+\frac{d^2}{d\beta_2^2})
\frac{d}{d s}
\log Z_{diff}(s) \mid_{\bar{\beta}=\bar{0}} ds   \ \ .
\label{eqn:Dtrace}
\end{equation}
The subscript $diff$ will be suppressed form now on.

We now consider the Lorentz gas obtained by infinitely
unfolding the Sinai billiard.
The relevant zeta function has a
BER approximation 
\EqRef{Zflow} 
\begin{equation}
Z_{BER}(s)=1-\langle e^{-s\Delta}e^{x_1\beta_1+x_2\beta_2} \rangle=
1-\langle e^{-s\Delta}\frac{1}{2\pi}\int_0^{2\pi}
e^{\Delta\beta cos(\theta)}d\theta  \rangle_{sos}  \ \ ,
\end{equation}
where we assumed isotropy (which is not really necessary cf.
refs. \cite{PDlyap,Ble92}), and $\Delta=\sqrt{x_1^2+x_2^2}$.
We only need to expand to
second order in $\beta$
\begin{equation}
Z_{BER}(s)=1-\langle e^{-s\Delta} \rangle -\frac{\beta^2}{4} 
\langle \Delta^2 e^{-s\Delta} \rangle \ldots
\end{equation}
We can now evaluate the averages by means of the distribution
$p_{\beta=0}$ discussed in section IIIb.
\begin{equation}
Z_{BER}(s)=1-
\left( \frac{s}{2R} +O(s^2 \log s) \right) -
\left( \frac{\beta^2}{4\pi^2R^2}
(\log s+O(1))  \right) 
\ldots   \label{eqn:ZDexpand}  \ \ .
\end{equation}
keeping only the leading terms in $R$. 
Inserting tis expression into
 \EqRef{Dtrace} we get as a result a diverging diffusion constant
\begin{equation}
D =\lim_{t \rightarrow} \frac{1}{2t} \frac{2}{\pi^2R}
\frac{1}{2\pi i}
\int_{-i\infty+\epsilon}^{i\infty+\epsilon} e^{st}\frac{-\log s +O(1)}{s^2} ds
=\frac{1}{\pi^2R}(\log(t)+O(1))  \ \ ,
\end{equation}
in agreement with the suggested exact
result \cite{Ble92} and with numerical simulations
\cite{DAcorr}.
This is one of the reasons why we expect that
$Z(s)=Z_{BER}(s)+O(s^2)$.

This anomalous diffusion behaviour of this Lorentz gas is a consequence of
the strongly intermittent properties of the Sinai billiard. 
Viewed as a Lorentz gas
the diverging diffusion constant is due to the 
unbounded horizon, the point particle
can make arbitrary long jumps between disks.

\csub{d. A short cut to the trace formula}

In investigations of the accuracy of the trace formula for billiards the
natural starting point is the boundary integral method \cite{Boas94,Prim95}.
Below we will, from this method, briefly derive the trace formula for the Sinai
billiard. The reason for this exercise is that it will help us
understand the
interplay between neutral and unstable orbits
in the next section.

According to the boundary integral method the eigenvalues of the problem
are those for which the following integral equation has a solution
\begin{equation}
u({\bf r}(s))=2\int_S \frac{\partial G}{\partial \hat{n}_s}
({\bf r}(s),{\bf r}(s'))u({\bf r}(s'))ds' \label{eqn:bim}  \ \ .
\end{equation}
The function $u({\bf r}(s))$ is related to the wave function 
according to
\begin{equation}
u({\bf r}(s))=\frac{\partial \Psi({\bf r}(s))}{\partial \hat{n}_s} \ \ .
\end{equation}
The Green function is arbitrary but we now adopt the idea of ref. \cite{Prim95}
and use the one-disk Green function, the integral \EqRef{bim}
should thus be performed only
along the square boundary.
The one disk Green function reads
\begin{equation}
G(r_1,r_2,\Delta\theta)=
\frac{i}{8}\sum_{\ell =-\infty}^{\infty}
\left(H_\ell^{-}(kr_1)+ S_\ell(kR) H_\ell^{+}(kr_1)\right)
H_\ell^{+}(kr_2) e^{i\ell(\Delta\theta)} \ \ ,
\end{equation}
where $H_\ell^{\pm}(z)$ are Hankel functions 
and $r_1$, $r_2$ and $\Delta \theta$ are explained in fig 3a. 
The phase shift function $S_\ell(kR)$ is defined by
\begin{equation}
S_\ell(kR) = -\frac{H_\ell^{-}(kR)}{H_\ell^{+}(kR)} \ \ .
\end{equation}

Using Poisson resummation we get
\begin{equation}
G(r_1,r_2,\Delta\theta)=
\sum_{m=-\infty}^{\infty}G^{(m)}(r_1,r_2,\Delta\theta) \ \ ,
\end{equation}
where
\begin{equation}
G^{(m)}(r_1,r_2,\Delta\theta)=\frac{i}{8}\int_{-\infty}^{\infty}
\left(H_\ell^{-}(kr_1)+ S_\ell(kR) H_\ell^{+}(kr_1)\right)
H_\ell^{+}(kr_2) e^{i\ell(\Delta\theta+2\pi m)}d\ell  \label{eqn:Gmint} \ \ .
\end{equation}
The standard semiclassical result are obtained by replacing all Hankel functions
by their Debye approximation and taking the integral approximation 
$G^{(m=0)}$ of the sum
(the higher terms in the Poisson resummation corresponds to creeping 
trajectories).
The integral is then performed by stationary phase.

We will now introduce an unnecessary approximation. The reason is just to make
the following reasoning more transparent. We will assume that $r_1,r_2 \gg R$.
This works well for small radii $R$ because the end points 
in $G(r_1,r_2,\Delta\theta)$ is kept on the square boundary.
In the semiclassical limit $k\rightarrow \infty$ the Green function is
\[
G(r_1,r_2,\Delta\theta)\approx
\]
\begin{equation}
 \left\{ \begin{array}{cc}
G_d(D)+G_d(R_1)\sqrt{4\pi kR \cos\phi }e^{-i\pi /4}G_d(R_2) & 
\Delta\theta< \arccos (R/r)+\arccos (R/r') \\
0 & \Delta\theta> \arccos (R/r)+\arccos (R/r') \end{array} \right.  
\label{eqn:circsmall} \ \ , \label{eqn:CircleDeb}
\end{equation}
expressing that in the lit region the Green function is a sum between the
direct trajectory and one reflected on the disk.
The lengths $D$, $R_1$ and $R_2$ are explained in fig 3a.
$G_d$ denote the asymptotic limit of the free (outgoing) Green function
\begin{equation}
G_d(L)=-\frac{i}{4}\sqrt{\frac{2}{\pi}}\frac{e^{i(kL-\pi/4)}}{\sqrt{kL}} \ \ .
\end{equation}
In \EqRef{circsmall} we have neglected the transitional behaviour around
$\Delta\theta\approx \arccos (R/r_1)+\arccos (R/r_2)$ which is referred to as
the penumbra in \cite{Prim95}, this is the subject of section IIIe.

We can write eq \EqRef{bim} symbolically as the matrix equation
\begin{equation}
({\bf I}-{\bf A}){\bf U}=0 \ \ ,
\end{equation}
having a solution only when $\det({\bf I}-{\bf A})=0$.
We can now express the
integrated density of states as
\begin{equation}
N(k)=\sum_{i=1}^{\infty} \Theta(k-k_i)=-\frac{1}{\pi}
\mbox{Im} \; \log \det ({\bf I}-{\bf A})
\end{equation}
\[=-\frac{1}{\pi} \mbox{Im} \; \mbox{tr} \log ({\bf I}-{\bf A})
=-\frac{1}{\pi} \mbox{Im} \; 
\sum_{n=1}^{\infty} \frac{1}{n}\mbox{tr} ({\bf A}^n) \ \ ,
\]
where
\begin{equation}
\mbox{tr} ({\bf A}^n)=2^n \int ds_1 \ldots ds_n 
\frac{\partial G}{\partial \hat{n}_{s_1}}({\bf r}(s_1),{\bf r}(s_2))
\ldots
\frac{\partial G}{\partial \hat{n}_{s_{n-1}}}({\bf r}(s_{n-1}),{\bf r}(s_n))
\frac{\partial G}{\partial \hat{n}_{s_n}}({\bf r}(s_n),{\bf r}(s_1)) 
\ \ . \label{eqn:traceAn}
\end{equation}

Performing the normal derivative on \EqRef{circsmall} we get (for large $k$) 
\begin{equation}
\frac{\partial G(r_1,r_2,\Delta\theta)}{\partial \hat{n}} \sim
(-ik)\cos (\theta_D) G_d(D)+
(-ik)\cos (\theta_R) G_d(R_1)\sqrt{4\pi kR cos\phi }e^{-i\pi /4}G_d(R_2) \ \ ,
\end{equation}
provided of course that we are in the lit region
$\Delta\theta< \arccos (R/r_1)+\arccos (R/r_2)$.
The angles $\theta_D$ and $\theta_R$ are explained in fig 3b.

The integrals in \EqRef{traceAn} will select classical orbits obeying the
classical reflection law. 
After performing the last integral in \EqRef{traceAn} only the 
periodic orbits will remain.
To compute the contribution from classical orbits
we will need to perform the following integrals 
by stationary phase analysis around a stationary point $s=s_0$
\begin{equation}
\int ds \; 
G_d(|{\bf r}'-{\bf r}(s)|)
2(-ik) \cos (\theta) G_d
(|{\bf r}(s)-{\bf r}''|)\sim
-G_d(|{\bf r}'-{\bf r}(s_0)|+|{\bf r}(s_0)-{\bf r}''|) \ \ .
\end{equation}

Periodic orbits come in two different types.
Orbits never bouncing on a disk are neutrally stable ({\em neutral} 
orbits for short)
and enter in one-parameter families.
Orbits bouncing at least once on a disk are unstable.
In the small $R$ limit the 
stability eigenvalues are approximately \cite{PDsmall}
\begin{equation}
\Lambda_p = \prod_{i=1}^{n_p} \frac{2l_i}{R\cos \phi_i} \ \ ,
\end{equation}
where $l_i$ is the 
traveling length between the hits on the disk and $\phi_i$ are
the scattering angles for these hits. 

The contribution to $tr({\bf A}^n)/n$ from a primitive unstable 
periodic $p$ with
length $L_p$ hitting the square boundary
$n_p$ times and traversed $r=n/n_p$ times is
\begin{equation}
\frac{1}{n}n_p\left( \prod_{i=1}^{n_p} \sqrt{4kR\cos (\phi_i )}
\sqrt{\frac{1}{8\pi k l_i}} \right )^r e^{ikrL_p}=
\frac{1}{r} \frac{1}{|\Lambda_p|^{r/2}} e^{ikr L_p+\chi_p r\pi}  
\label{eqn:singutz} \ \ .
\end{equation}
The factor $n_p$ in front accounts for the number of periodic points
on the square boundary and $\chi_p$ counts the number of bounces along the 
orbit. 
The result above was obtained in the small $R$ limit. For finite $R$ the factor
$|\Lambda_p|^{r/2}$ should be replaced by $|\Lambda_p|^{r/2}(1-1/\Lambda_p^r)$.
The unstable orbits give rise to Gutzwiller's trace formula which can be 
recasted
to the Gutzwiller-Voros zeta function.
The neglect of the factor $(1-1/\Lambda_p^r)$ is equivalent to the 
neglect of the higher
$m$ factors in \EqRef{GVzeta} which is not our primary concern.

A primitive neutral orbit (or rather a one-parameter family of neutral orbits) 
of length $L_q$ and geometric width $D_q$ and repetition number $r=n/n_q$
give
the following contribution to 
$tr({\bf A}^n)/n$
\begin{equation}
\frac{1}{n}n_q D_q 
\left(  -\frac{i}{4}\sqrt{\frac{2}{\pi}}\frac{1}{\sqrt{kL_q r}} \right)
e^{i(kL_q r-\pi/4)+\chi_q r\pi}  \label{eqn:sinBT} \ \ .
\end{equation}
Now only $n-1$ integrals are computed by stationary phase, the last 
integral gives rise
to the factor $D_q$. 

The important lesson from this section 
is that the dichotomy between unstable and neutral orbits in the trace formula
is a direct consequence of eq. \EqRef{CircleDeb}. 
In the next section we will see how the neglect of the quantum fuzziness in
the twilight zone between the reflection and direct region in the 
Green function
\EqRef{CircleDeb} allows singularities to enter the Gutzwiller-Voros 
zeta function.

\csub{e. How singularities sneaks into the Gutzwiller-Voros zeta function}

We will consider scatterings in extreme forward angles so only the second
term in the integral \EqRef{Gmint} (m=0) contributes.
This means that we only need to evaluate the integral
\begin{equation}
G^{(0)}(r_1,r_2,\Delta\theta)=
\frac{i}{8}\int_{-\infty}^{\infty}S_\ell(kR) H_\ell^{+}(kr_1)
H_\ell^{+}(kr_2) e^{i\ell \Delta\theta}d\ell  \label{eqn:Jdef} \ \ ,
\end{equation}
where $0\ll \Delta\theta<\pi$. 
We can still safely use the Debye approximation for the 
Hankel functions $H^+_\ell(kr)$ and
$H^+_\ell(kr')$  
\begin{equation}
H_\ell^{+}(z)\sim \sqrt{\frac{2}{\pi\sqrt{z^2-\ell^2}}}
e^{i[\sqrt{z^2-\ell^2}-\ell\; \arccos (\ell/z)-\pi/4]} \ \ ,
\end{equation}
because the end points stay on the square boundary. However,
the phase shift function needs a more careful analysis. 
The phase shift function $S_\ell (kR)$ is of unit modulus
and we call the phase $\gamma(kR,\ell )$:
\begin{equation}
S_\ell(kR) = 
-\frac{H_\ell^{-}(kR)}{H_\ell^{+}(kR)}\equiv e^{i\gamma(kR,\ell)} \ \ .
\end{equation}
The Green function now reads
\begin{equation}
G^{(0)}(r_1,r_2,\Delta\theta)=\int_{-\infty}^{\infty}
A(r_1,r_2)
e^{i(\ell\Delta\theta - \Psi(\ell))}
d\ell  \ \ ,
\end{equation}
where $A(r_1,r_2)$, coming from the Hankel functions $H_\ell^+ (kr_1)$
and $H_\ell^- (kr_2)$,
is a slowly varying amplitude so that the integral is
in principle determined by the phase function
\begin{equation}
\Psi(\ell)=-\gamma(kR,\ell)-[\sqrt{(kr)^2-\ell^2}-\ell\; \arccos (\ell/(kr))]
-[\sqrt{(kr')^2-\ell^2}-\ell\; \arccos (\ell/(kr'))]+\pi/2 \ \ ,
\end{equation}
and the stationary phase condition will simply read
$\Psi_{\ell}(\ell)=\Delta\theta$ (subscripts denote
differentiation).

We will now investigate
the phase of $S_\ell(kR)$ in detail.
To this end we will use the uniform approximation for Hankel functions relating
the phase of Hankel functions 
to the phase of Airy functions $\theta$ according to \cite{AS}
\begin{equation}
S_\ell(kR)=-\frac{\mbox{Ai}(-x)+i\mbox{Bi}(-x)}{\mbox{Ai}(-x)-i\mbox{Bi}(-x)}
\equiv e^{2\theta(x)-\pi}\equiv{e^{i\gamma(\ell,kR)}}
\end{equation}
where
\begin{equation}
x(kR,\ell)=\left\{ \begin{array}{cc}
[\frac{3}{2}(\sqrt{(kR)^2-\ell^2}-\ell \; \arccos(\ell/(kR)))]^{2/3} 
& \ell<kR\\
-[\frac{3}{2}(\ell\cdot\log (\frac{\ell+\sqrt{\ell^2-(kR)^2}}{kR})-
\sqrt{\ell^2-(kR)^2}]^{2/3} & \ell>kR \end{array}\right.  \ \ .
\end{equation}
If $|kR/\ell-1| \ll 1$ we can use
\begin{equation}
x=(\frac{2}{\ell})^{1/3}(kR-\ell)  \label{eqn:xappr} \ \ .
\end{equation}
Using this expression for $x$ one gets the so called transition region
approximation.
This approximation fails to yield the Debye approximation as its 
asymptotic limit
but the nice thing is that there is a considerable overlap 
because whenever $x\ll \ell^{2/3}$ we can use the transition region 
approximation and when
$x \gg 1$ we can use Debye. (The relevant $\ell$ for our consideration 
scales as
$\ell \sim k$).

$\theta(x)$ is
a complicated function but one has the following asymptotic 
expressions \cite{AS}
\begin{eqnarray}
\theta(x)=\pi/3-\frac{3^{4/3}\Gamma(2/3)^2}{4\pi}x-
\frac{3^{5/3}\Gamma(2/3)^3}{8\pi
\Gamma(1/3)}x^2+O(x^4) & x \rightarrow \pm 0\\
\theta(x) \sim 
\pi/2 -\frac{1}{2} e^{-\frac{4}{3}(-x)^{3/2}}(1+O(1/(-x)^{3/2})) &
x\rightarrow-\infty  \ \ ,  \\
\theta(x) =\frac{\pi}{4}-\frac{2}{3}x^{3/2}+O(x^{-3/2}) & x \rightarrow +\infty
\end{eqnarray}
so we obtain the Debye approximation when $x \rightarrow \infty$
as expected.

In fig 4 we plot the function $\Psi_\ell (\ell)$ for some arbitrary choose
values of $r_1/R=r_2/R=3$ and $kR=10$ together with its Debye approximation.
We have marked some critical values of $\Psi_\ell (\ell)$ in the plot:
\begin{itemize}
\item $\Delta\theta_{cl}=\arccos(R/r_1)+\arccos(R/r_2)$ 
denotes the limit of the classically illuminated
region. 
\item $\Delta\theta_d$ is the limit of the standard semiclassical
result on
the direct side. It scales with $kR$ as 
$\Delta\theta_d=\Delta\theta_{cl}-O((kR)^{-2/3})$.
\item $\Delta\theta_r$ is the corresponding limit on
the reflection side. It scales as 
$\Delta\theta_r=\Delta\theta_{cl}-O((kR)^{-1/3})$.
\item The maximum of the curve occurs at 
$\Delta\theta_{max}=\Delta\theta_{cl}-O((kR)^{-2/3}\log(kR)^{2/3})$ and at
$\ell=\ell_{max}$. The second derivative of $\Psi_{\ell}$
at the maximum is
$\Psi_{\ell \ell \ell}(\ell_{max})=O( (kR)^{-4/3}(\log kR)^{1/3})$.
\end{itemize}

If $\Delta \theta=\Delta\theta_{max}+\delta$ 
and $\delta$ is small,
the Green
function is still non zero but cannot be evaluated by stationary phase.
This means that we cannot in an unambiguous way divide the Green function
into one part associated with
the direct ray and one with the reflected, as in \EqRef{CircleDeb},
 and something in the semiclassical 
description is definitely lost.

Close to the maximum the result can be approximated by the integral
\begin{equation}
\int_{- \infty}^{\infty} e^{i(\delta \ell +\frac{1}{6} |\Psi_{\ell \ell
\ell}|(\ell - \ell_{\max})^3)}d \ell
=2\pi |\Psi_{\ell \ell \ell}/2|^{-1/3} 
\mbox{Ai}(|\Psi_{\ell \ell \ell}/2|^{-1/3} \delta)
\end{equation}
From this we can get an estimate of the critical value of $\delta$ when
stationary phase approximation ceases to be valid. This happens when the
argument of the
Airy function is $O(1)$. We can define a critical
$\delta$ by
\begin{equation}
\delta_{crit}|\Psi_{\ell \ell \ell}(\ell_{max})|=O(1) \ \ ,
\end{equation}
and we find that $\delta_{crit}$ are of the order
\begin{equation}
\delta_{crit}= O((kR)^{-4/9} (\log kR)^{1/9}) \ \ .
\end{equation}
We only expect the Airy function approximation to work well close to the 
maximum
since the curve $\Psi_\ell$ is very skew and badly approximated by a parabola.

In ref. \cite{Prim95} the authors choose to approximate
the direct side $\ell>k$ by a Fresnel integral (this assuming 
constant slope of $\Psi_\ell$)
but the treatment on the reflections side is incomplete.

Let us now consider the family $q_n$ of periodic orbits
indicated  in fig 1. We see that if $n$ is big enough
the unstable periodic orbit will interfere
with the neutral orbit and for large enough $n$ the 
corresponding terms in the Gutzwiller
trace formula are suppressed.
More precisely, it is the stationary points closest to the disk (cf.
the one marked with a cross in fig 1) that 
suffers most of 
interference with the edge
of the corresponding family of neutral orbits.
In the same way as
there is no longer an unambiguous division between a direct and a
reflected ray there is no longer
a corresponding division between neutral and
unstable orbits in the trace formula, cf. section IIId.
It is the neglect of this effect that allows the branch point singularity
in the Gutzwiller-Voros zeta function.

The critical $n=n_{crit}$ when this interference problem gets crucial
scales as 
\begin{equation}
n_{crit} \sim (kR)^{2/3}  \  \  .
\end{equation}
which is
the same as the critical threshold of ref. \cite{Prim95} above which
the majority of periodic orbits suffer from 'diffraction' effects.

It is this type of
entanglement which makes semiclassical quantization
of systems with mixed phase space so difficult.
Our findings are very similar to those of ref.\cite{Tann96}
concerning the Stadium billiard.


\vspace{1cm}
I would like to thank Karl-Erik Thylwe for discussions on circle 
Green functions. 
To result in section IIIe
will appear in a joint publication.
This work was supported by the Swedish Natural Science
Research Council (NFR) under contract no. F-FU 06420-303.

\newcommand{\PR}[1]{{Phys.\ Rep.}\/ {\bf #1}}
\newcommand{\PRL}[1]{{Phys.\ Rev.\ Lett.}\/ {\bf #1}}
\newcommand{\PRA}[1]{{Phys.\ Rev.\ A}\/ {\bf #1}}
\newcommand{\PRD}[1]{{Phys.\ Rev.\ D}\/ {\bf #1}}
\newcommand{\PRE}[1]{{Phys.\ Rev.\ E}\/ {\bf #1}}
\newcommand{\JPA}[1]{{J.\ Phys.\ A}\/ {\bf #1}}
\newcommand{\JPB}[1]{{J.\ Phys.\ B}\/ {\bf #1}}
\newcommand{\JCP}[1]{{J.\ Chem.\ Phys.}\/ {\bf #1}}
\newcommand{\JPC}[1]{{J.\ Phys.\ Chem.}\/ {\bf #1}}
\newcommand{\JMP}[1]{{J.\ Math.\ Phys.}\/ {\bf #1}}
\newcommand{\JSP}[1]{{J.\ Stat.\ Phys.}\/ {\bf #1}}
\newcommand{\AP}[1]{{Ann.\ Phys.}\/ {\bf #1}}
\newcommand{\PLB}[1]{{Phys.\ Lett.\ B}\/ {\bf #1}}
\newcommand{\PLA}[1]{{Phys.\ Lett.\ A}\/ {\bf #1}}
\newcommand{\PD}[1]{{Physica D}\/ {\bf #1}}
\newcommand{\NPB}[1]{{Nucl.\ Phys.\ B}\/ {\bf #1}}
\newcommand{\INCB}[1]{{Il Nuov.\ Cim.\ B}\/ {\bf #1}}
\newcommand{\JETP}[1]{{Sov.\ Phys.\ JETP}\/ {\bf #1}}
\newcommand{\JETPL}[1]{{JETP Lett.\ }\/ {\bf #1}}
\newcommand{\RMS}[1]{{Russ.\ Math.\ Surv.}\/ {\bf #1}}
\newcommand{\USSR}[1]{{Math.\ USSR.\ Sb.}\/ {\bf #1}}
\newcommand{\PST}[1]{{Phys.\ Scripta T}\/ {\bf #1}}
\newcommand{\CM}[1]{{Cont.\ Math.}\/ {\bf #1}}
\newcommand{\JMPA}[1]{{J.\ Math.\ Pure Appl.}\/ {\bf #1}}
\newcommand{\CMP}[1]{{Comm.\ Math.\ Phys.}\/ {\bf #1}}
\newcommand{\PRS}[1]{{Proc.\ R.\ Soc. Lond.\ A}\/ {\bf #1}}
%


\begin{center}

\end{center}

\newpage

\csec{Figure captions}

\FigCap{1}{The Sinai billiard (upper left corner) is the unit cell 
of the associated
Lorentz gas. Two transparent direction are indicated (shaded areas).
The family $q_n$ of periodic orbits accumulating towards the horizontal
transparent direction is also indicated.}

\FigCap{2}{The intermittent map discussed in section IIIb.
The sequence $\{ c_n\}$ is are defined as the inverse images of the 
critical point
as indicated.}

\FigCap{3}{The semiclassical limit of the 
Green function between two points on the border can be divied into
two parts. One associated with the direct trajectory and one that is reflected
on the disk.}

\FigCap{4}{The function $\Psi_\ell$ discussed in section IIIe together with
its Debye approximation.}

\end{document}